\documentclass[prl,twocolumn,showpacs,preprintnumbers,amsmath,amssymb,superscriptaddress]{revtex4-1}

\usepackage{amsmath,amssymb,amsfonts} 	
\usepackage{graphicx}
\usepackage[latin1]{inputenc}
\usepackage{subfigure}

\renewcommand{\[}{\begin{equation}}
\renewcommand{\]}{\end{equation}}
\def\bea{\begin{eqnarray}}
\def\eea{\end{eqnarray}}
\def\nn{\nonumber\\}

\newcommand{\intrc}{\int_{\rm cell}\!\!\!\! d{\bf r} \;}
\newcommand{\emi}[1]{{\rm e}^{-i #1}}
\newcommand{\ei}[1]{{\rm e}^{i #1}}

\newcommand{\dk}{[d \, {\rm k}]}
\newcommand{\intk}{\int_{\rm BZ} \!\!\!\! [d \, {\rm k}] \;}

\newcommand{\CQ}{{\cal Q}}
\newcommand{\CP}{{\cal P}}

\newcommand{\q}{{\bf k}}
\renewcommand{\k}{{\bf k}}

\newcommand{\vc}{V_{\rm cell}}
\newcommand{\CH}{{\cal H}}
\newcommand{\dkk}{\partial_{{\bf k}}}

\newcommand{\trv}{\mbox{Tr$_V$}}
\newcommand{\traa}{\mbox{Tr$_A$}}

\renewcommand{\r}{{\bf r}}

\newcommand{\da}{\partial_{k_\alpha}\!}
\newcommand{\db}{\partial_{k_\beta}\!}

\newcommand{\equ}[1]{Eq.~(\ref{#1})}
\newcommand{\eqs}[2]{Eqs.~(\ref{#1}) and (\ref{#2})}

\def\bra#1{\langle#1\vert}
\def\ket#1{\vert#1\rangle}
\def\ev#1{\langle#1\rangle}
\def\me#1#2#3{\langle#1| \, #2 \, |#3\rangle}
\def\runtime{(\the\time)\qquad\the\month/\the\day/\the\year}
\def\today
 {\count10=\year\advance\count10 by -2000 \number\day--\ifcase
  \month \or Jan\or Feb\or Mar\or Apr\or May\or Jun\or
             Jul\or Aug\or Sep\or Oct\or Nov\or Dec\fi--\number\count10}

\def\hour{\count10=\time\count11=\count10
\divide\count10 by 60 \count12=\count10
\multiply\count12 by 60 \advance\count11 by -\count12\count12=0
\number\count10 :\ifnum\count11 < 10 \number\count12\fi\number\count11}

\begin{document}

\title{Locality of the anomalous Hall conductivity}

\author{Antimo Marrazzo}
\email{antimo.marrazzo@epfl.ch}
\affiliation{Theory and Simulation of Materials (THEOS), \'Ecole Polytechnique F\'ed\'erale de Lausanne, CH-1015 Lausanne, Switzerland}

\author{Raffaele Resta}
\email{resta@democritos.it}
\affiliation{Dipartimento di Fisica, Universit{\`a} di Trieste, 34127 Trieste, Italy}
\affiliation{Consiglio Nazionale delle Ricerche (CNR), Istituto Officina dei Materiali (IOM) DEMOCRITOS, 34136 Trieste, Italy}
\affiliation{Donostia International Physics Center, 20018 San Sebasti{\'a}n, Spain}

\date{\today}

\begin{abstract} The geometrical intrinsic contribution to the anomalous Hall conductivity (AHC) of a metal is commonly expressed as a reciprocal-space integral: as such, it only addresses unbounded and macroscopically homogeneous samples. Here we show that the geometrical AHC has an equivalent expression as a local property. We define a ``geometrical marker'' which actually probes the AHC in inhomogeneous systems (e.g. heterojunctions), as well as in bounded samples. The marker may even include extrinsic contributions of geometrical nature.
\end{abstract}

\date{run through \LaTeX\ on \today\ at \hour}

\maketitle\bigskip \bigskip
The Hall conductivity is ``anomalous'' whenever it is nonzero in absence of an applied magnetic field. The phenomenon requires the absence of time-reversal symmetry: it was discovered by Hall himself in 1881 in ferromagnetic metals. The possibility of observing anomalous Hall conductivity (AHC) in insulators was pointed out in 1988 by Haldane, who proposed a model Hamiltonian where the AHC is nonzero and quantized \cite{Haldane88}; the AHC value is determined by the topology of the electronic ground state. In metals extrinsic mechanisms are essential to make the longitudinal dc conductivity finite. In absence of time-reversal symmetry extrinsic mechanisms contribute to the AHC as well: these go under the name of side-jump and skew-scattering \cite{Nagaosa10}. Since the early 2000s \cite{Jungwirth02,Onoda02} it became clear that an {\it intrinsic} effect, only dependent on the ground wavefunction of the pristine crystal, provides an important additional contribution to the AHC. The latter contribution is geometrical in nature; its expression is the nonquantized version of the corresponding formula for insulators. In this work we only address the geometrical/topological AHC in metals/insulators, which is customarily expressed as the Fermi-volume integral of the Berry curvature, \equ{ss} below. The standard approach requires an unbounded crystalline sample where the orbitals have the Bloch form; this was extended in Ref. \cite{rap149} to ``dirty'' metals in a supercell framework, where the geometric contribution includes some extrinsic effects.

Recent work has demonstrated the {\it locality} of AHC, although in the insulating case only \cite{rap146}. One does not require lattice periodicity and reciprocal-space paraphernalia: the AHC can be defined and computed for bounded samples and/or for macroscopically inhomogeneous systems (e.g. heterojunctions). The extension to the metallic case is not obvious, since one of the reasons for the locality of the (quantized) AHC is the $\r$-space exponential decay of the one-body density matrix in insulators (``nearsightedness'' \cite{Kohn96}). In metals instead such decay is only power-law, which hints to a possibly different behavior. Furthermore the Hall current in insulators may only flow in the edge region, while in metals the current flows through the bulk of the sample as well. Our major result is that even in metals the AHC is a local property: for a bounded sample it can be expressed in terms of the one-body density matrix, evaluated in the sample bulk. We show this by means of simulations on model two-dimensional bounded samples (flakes) by adopting the---by now famous---Haldane model Hamiltonian \cite{Haldane88}.

We start reviewing the well established expression for the intrinsic AHC in crystalline insulators and metals; as said above, the formula is basically the same in both cases. It yields a quantized (topological) AHC in insulators, and a nonquantized (geometrical) AHC in metals.
We get rid of trivial factors of two throughout, thus addressing ``spinless electrons''. The crystalline orbitals have the Bloch form $\ket{\psi_{j \k}} = \ei{\k \cdot \r} \ket{u_{j \k}}$; here they are normalized to one over the unit cell. The periodic orbitals $\ket{u_{j \k}}$ are eigenstates of $\CH_\k = \emi{\k \cdot \r} \CH \ei{\k \cdot \r}$, and we choose a gauge which makes them smooth in the whole Brillouin zone (BZ).

 The ground-state projector can be written as 
\bea \me{\r}{\CP}{\r'} &=& \vc \intk \ei{\k \cdot (\r -\r')} \me{\r}{\CP_\k}{\r'} \label{CP1} \\ \CP_{\k} &=&  \sum_j \theta(\mu - \epsilon_{j\q}) \ket{u_{j\q}} \bra{u_{j\q}} ,\label{CP2} \eea  where $\mu$ is the Fermi level, the BZ integration is over $\dk = d\k/(2\pi)^d$ ($d$ is the dimension, either 2 or 3), and $\vc$ is the cell volume (area for $d=2$). We will also need the complementary projectors $\CQ = {\cal I} - \CP $ and $\CQ_\k = {\cal I} - \CP_\k $.

The Berry curvature of the occupied manifold  is
\bea \Omega_{\alpha\beta}(\k) &=& - 2\, \mbox{Im } \sum_j  \theta(\mu - \epsilon_{j\q}) \ev{\da u_{j\q}| \db\ u_{j\q}}  \label{curva} \\ &=& - 2\, \mbox{Im } \sum_j  \theta(\mu - \epsilon_{j\q}) \me{\da u_{j\q}}{\CQ_\k}{ \db u_{j\q}} ; \nonumber \eea it is smooth over the whole BZ is for insulators, while it is piecewise smooth (and integrable) for metals.

The intrinsic AHC contribution is \[ \sigma_{\alpha\beta} = - \frac{e^2}{\hbar} \intk \Omega_{\alpha\beta}(\k) , \label{ss} \]
where the BZ integral is actually a Fermi-volume integral in the metallic case, owing to the $\theta$ function in \equ{CP2}.
\equ{ss} as it stands holds for both $d=2$ and $d=3$; we further notice that $\sigma_{\alpha\beta}$---when expressed in $e^2/h$ units (a.k.a. klitzing$^{-1}$)---is dimensionless for $d=2$, while it has the dimensions of an inverse length for $d=3$. 

The position operator $\r$ is notoriously ill defined within periodic boundary conditions \cite{rap100}; nonetheless its off-diagonal elements over the $\ket{\psi_{j\k}}$ and $\ket{u_{j\k}}$ are well defined. Exploiting some results from linear-response theory \cite{Baroni01}, one may prove that \[ \CQ_\k \r \ket{u_{j\k}} = i \CQ_\k \ket{\dkk u_{j\k}} \] whenever $j$ labels an occupied state at the given $\k$. We may thus write the Berry curvature as a trace; \[ \Omega_{\alpha\beta}(\k) = - 2\, \mbox{Im } \mbox{Tr } \{ \CP_\k r_\alpha \CQ_\k  r_\beta \} . \] Using then the definitions of $\CP$ and $\CP_\k$ (and their complementary), \eqs{CP1}{CP2}, it is easy to prove the identity
\[ \frac{1}{\vc} \intrc \me{\r}{\CP r_\alpha \CQ  r_\beta}{\r} = \intk \mbox{Tr } \{ \CP_\k r_\alpha \CQ_\k  r_\beta \} . \label{iden} \] This identity is known since a few years \cite{Bellissard94,rap118,Kitaev06,Prodan09,Prodan10,rap146} for the insulating case---and for the insulating case {\it only}. We stress that the alternate proof provided here applies to the metallic case as well. The l.h.s. of \equ{iden} has two outstanding virtues: (i) it is expressed directly in the Sch\"rodinger representation, making no reference to reciprocal space; and (ii) can be adopted as such for supercells of arbitrarily large size, thus extending the concept of geometrical AHC to disordered systems, such as alloys, as well as ``dirty'' metals and insulators. We thus recast \eqs{ss}{iden} in the compact form \bea \sigma_{\alpha\beta} &=& \frac{2 e^2}{\hbar} \mbox{Im } \trv  \{ \CP r_\alpha \CQ  r_\beta \} \nn &=& - \frac{2 e^2}{\hbar} \mbox{Im } \trv  \{ \CP \, [r_\alpha,\CP] \,   [r_\beta,\CP ] \} , \label{trv} \eea where ``\trv'' means trace per unit volume/area. The two expressions in \equ{trv} are formally equivalent; the second one, being a $\CP$-only formula, is more suited to numerical implementations.

We pause at this point to make contact with Ref. \cite{rap149}, where a supercell approach to dirty metals was actually proposed: in retrospective, the approach of Ref. \cite{rap149} is equivalent to evaluating \equ{trv} over the folded BZ of the superlattice. Indeed \equ{trv}, when applied to a dirty metal, combines the nominally intrinsic contribution---as defined for the clean metal---to some extrinsic contributions of geometrical nature. Following the arguments of Ref. \cite{rap149} we argue here that \equ{trv} may yield the sum of the intrinsic and side-jump contributions to the AHC, while instead it may not include the skew scattering \cite{Nagaosa10}.

Our major result so far, \equ{trv}, applies to either insulators or metals, either crystalline or disordered, but it has only been proved for an unbounded and macroscopically homogeneous system within periodic boundary conditions. The next issue is whether one may adopt \equ{trv} {\it locally}, in order to address inhomogeneous systems (e.g. heterojunctions) or even bounded samples (e.g. crystallites). 

The locality of the AHC was investigated in Ref. \cite{rap146}, where it was shown---for the insulating case only---that the topological AHC can indeed be evaluated from \equ{trv} for bounded and/or macroscopically inhomogeneous systems. The concept of ``topological marker'' was proposed therein; in the following we are going to show that \equ{trv} yields an analogous ``geometrical marker'', effective in the metallic case as well. The very important feature pointed out by Ref. \cite{rap146} is that---when a bounded sample is addressed---the trace per unit volume has to be evaluated using only some inner region of the sample, and {\it not} the whole sample. If the bounded system is a crystallite, one evaluates e.g. the l.h.s. of \equ{iden} over its central cell; in the large-crystallite limit one recovers the bulk value of the AHC. In all the cases dealt with in Ref. \cite{rap146} the convergence with size proved to be very fast: this was attributed to the exponential decay of the one-body density matrix in insulators (``nearsightedness'' \cite{Kohn96}), as already said in the Introduction. For the metallic case we are going to explore in the following an uncharted territory by means of case-study simulations.

\begin{figure}[t]
\centering
\includegraphics[width=\columnwidth]{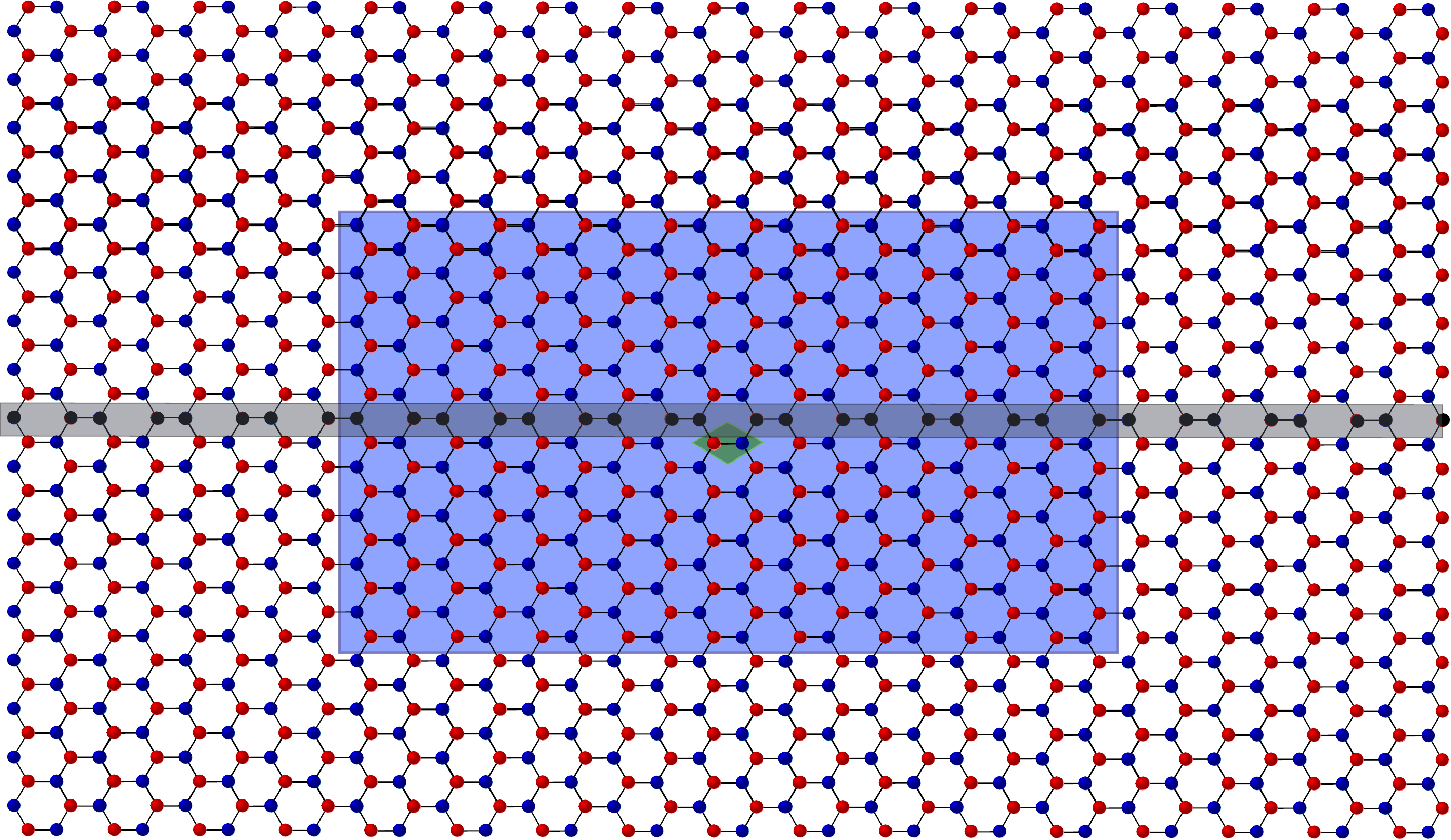}
\caption{(color online). A typical ``Haldanium'' flake. We have considered flakes with up to 10506 sites, all with the same aspect ratio; the one shown here has 1190 sites. In order to probe the AHC locality we evaluate the trace per unit area either on the central cell (two sites) or on the ``bulk'' region (1/4 of the sites). The grey horizontal line (black dots) highlights the sites chosen for drawing Fig. \ref{fig:hetero}.}
\label{fig:flake} \end{figure}

The paradigmatic model for investigating issues of the present kind is the one proposed by Haldane in 1988 \cite{Haldane88}. It is a tight-binding $2d$ Hamiltonian on a honeycomb lattice with onsite energies $\pm \Delta$, first neighbor hopping $t_1$, and second neighbor hopping $t_2 = |t_2| \ei{\phi}$, which provides time-reversal symmetry breaking. The model is insulating at half filling and metallic at any other filling. Our bounded samples are rectangular Haldanium flakes such as the one shown in Fig. \ref{fig:flake}; the corresponding simulations for lattice-periodical samples, with Bloch orbitals, are performed by means of the PythTB code \cite{PythTB}. Oscillations as a function of the flake size occur in the metallic case; as customary, we adopt a regularizing ``smearing'' technique.

\begin{figure}[t]
\centering
\includegraphics[width=\columnwidth]{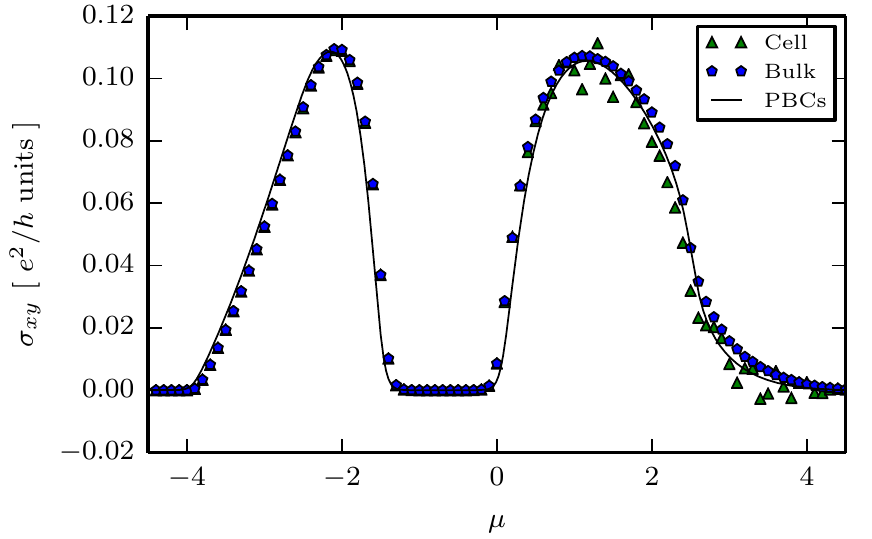}
\includegraphics[width=\columnwidth]{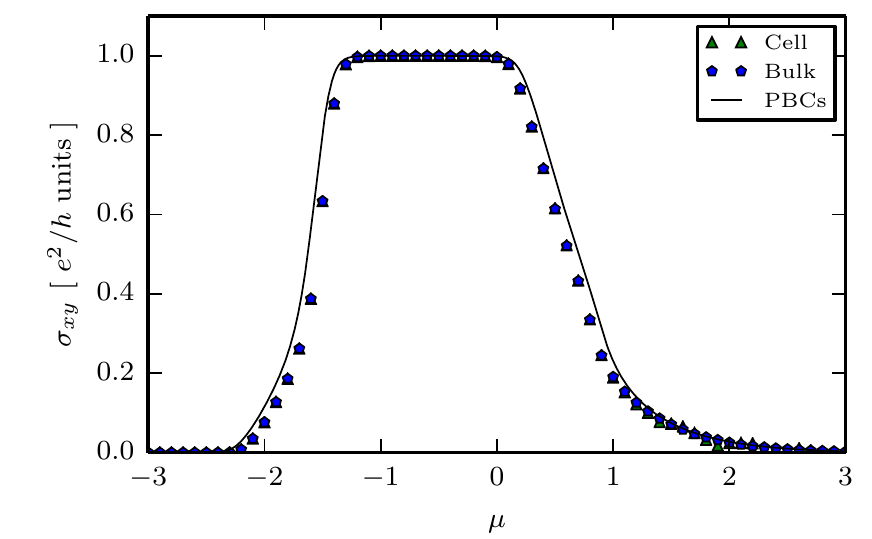}
\caption{AHC as a function of the Fermi level $\mu$ for a 3422-site flake. Top: trivial insulator when $\mu$ is in the gap; Bottom: topological insulator ($C_1 = -1$) when $\mu$ is in the gap. See text about labels: Cell, Bulk, and PBCs. All calculations adopt a ``smearing'' $s = 0.05$.
}
\label{fig:ahe} \end{figure}

\begin{figure}[t]
\centering
\includegraphics[width=\columnwidth]{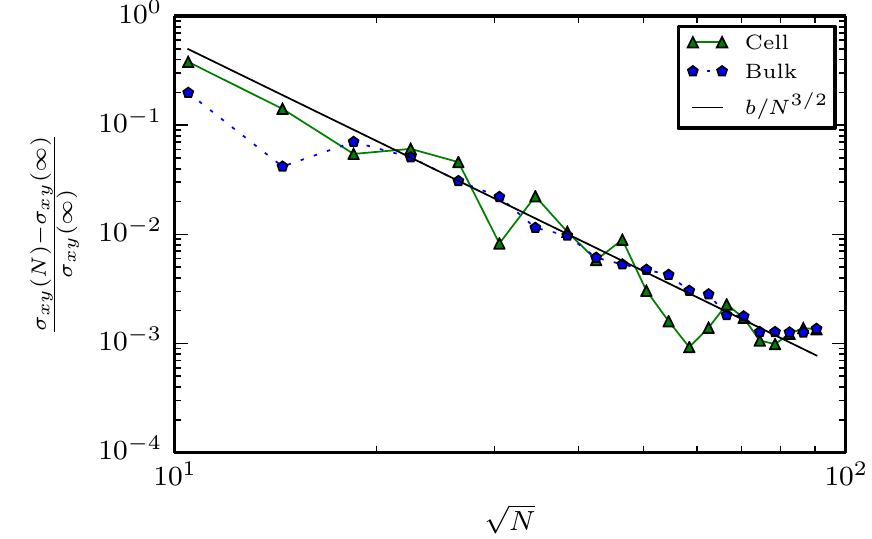}
\includegraphics[width=\columnwidth]{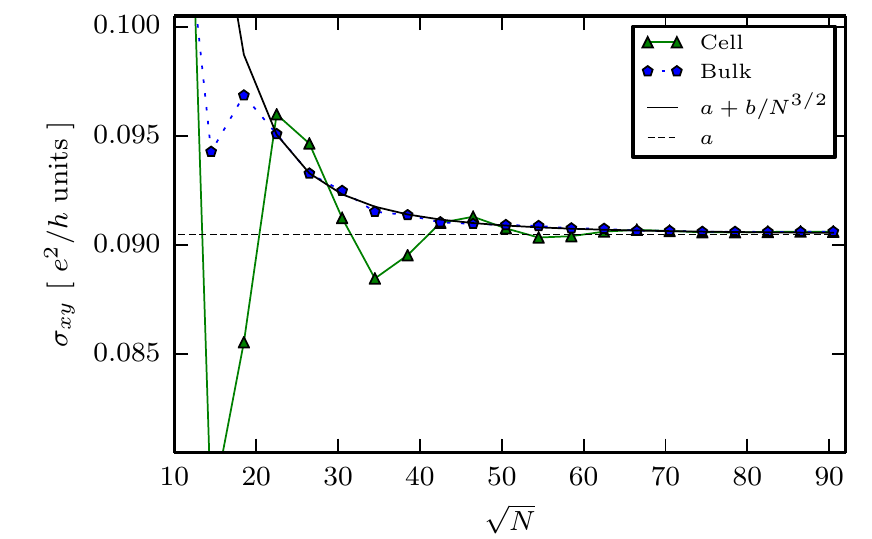}
\caption{Convergence of AHC evaluated locally as a function of the flake size. Parameters as in the top panel of Fig. \ref{fig:ahe}, and $\mu = -2.5$. The quantity $\sigma_{xy}(\infty)$ is obtained via extrapolation in the large flake limit. A smearing $s = 0.05$ is adopted.
}
\label{fig:conv} \end{figure}

In Fig. \ref{fig:ahe} we plot---as a function of the Fermi level $\mu$---the dimensionless quantity \[ - 4 \pi  \,\mbox{Im } \traa  \{ \CP \, [r_\alpha,\CP] \,   [r_\beta,\CP ]\} = \frac{h}{e^2} \sigma_{xy} , \label{tra}  \] where ``$\traa$'' means trace per unit area. The quantity in \equ{tra} equals minus the Chern number $C_1$ in the quantized insulating case \cite{sign}: nonzero $C_1$ reveals the nontrivial (topological) nature of the insulating ground state. Each panel displays the trace per unit area, \equ{tra}, evaluated in three different ways: over the central two sites (labeled ``Cell''), evaluated over 1/4 of the sites (labeled ``Bulk''), and evaluated as the usual integral of the Berry curvature for an unbounded sample (labeled ``PBCs''). The plots show that averaging over the bulk region provides a better convergence.
The two plots refer to two different sets of parameters: in both cases we set $t_1=1$ and $\phi= 0.25$, while $\Delta = 2$ the for top plot and $\Delta = 1/3$ for the bottom plot. It is perspicuous from the figure that when $\mu$ is in the gap region the former choice yields a trivial insulator, and the latter a topological one ($C_1 = -1$). 

Fig. \ref{fig:ahe} proves our major claim: the geometrical/topological AHC, for both metals and insulators, is indeed a {\it local} property of the electronic ground state and can be evaluated for a bounded sample, where the orbitals are square-integrable and the concept of reciprocal space does not make any sense. What differentiates insulators from metals is only the kind of convergence with the system size: exponential in the former case, power-law in the latter.
We show a typical convergence study in Fig. \ref{fig:conv}, where we have chosen a metallic flake with $\mu = -2.5$ and the Hamiltonian for which the corresponding insulator is trivial: top panel of Fig. \ref{fig:ahe}. As for the previous figure, averaging over the bulk region provides a better convergence than taking the trace on the central two-site cell. Interpolations in both panels clearly show that the AHC convergence to the bulk value is of the order $L^{-3}$, where $L$ is the linear size of the flake.

We have focused so far on the imaginary part of \equ{iden} only; here we briefly address the real part of the l.h.s. as well. It is known that---in the large-system limit---it converges to a finite value in insulators, while it diverges in metals \cite{rap_a31}. Simulations and heuristic arguments altogether suggest that the metallic divergence is of the order $L$ in any dimension: $d=1$, 2 or 3 \cite{rap143,Bendazzoli12,tesi}. Reasoning by analogy we conjecture that even the AHC convergence is likely to be dimension-independent, i.e. $L^{-3}$ for both $d$=2 and 3.

\begin{figure}[t]
\centering
\includegraphics[width=\columnwidth]{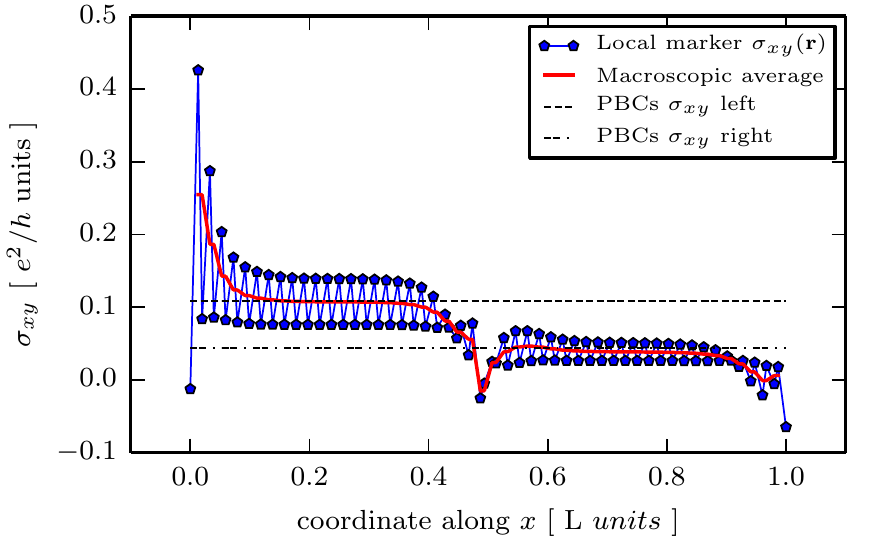}
\caption{Local AHC for an heterojunction, where the left- and right-half of the flake are two different metals (see text). For this calculation the flake has 10506 sites; our local function is shown on a line of 102 sites (grey area in Fig. \ref{fig:flake}). The two horizontal lines (labeled ``PBCs'') show the corresponding Berry-curvature calculations.
}
\label{fig:hetero} \end{figure}

Finally, we demonstrate the AHC locality on the case study of an heterojunction, where the two metals in the left- and right-half of the rectangular flake are different. In both regions we adopt an Hamiltonian like the one in the top panel of Fig.\ref{fig:ahe}, and we set $\mu=-2.2$; then we add a constant onsite energy equal to $+1$ on the right. 
We show in Fig. \ref{fig:hetero} the function $- 4 \pi\; \mbox{Im } \me{\r}{\CP \, [r_\alpha,\CP] \,   [r_\beta,\CP ]}{\r}$, normalized over the atomic area, along a direction normal to the heterojunction: the grey central line in Fig. \ref{fig:flake}. The macroscopic average (i.e. two-site average) of our function in the bulk of each region clearly yields the corresponding AHC, closely matching two standard $\k$-space calculations for the same 2$d$ metals, where $\mu= -2.2$ and $\mu= -3.2$, respectively.

\bigskip

In conclusion, we have shown that the geometrical contribution to the AHC does not need a reciprocal-space approach to be defined and computed. We have provided a ``geometrical marker'' defined in $\r$ space, \equ{trv}.  The marker probes the electron distribution {\it locally}, and may therefore address inhomogeneous systems (e.g. heterojunctions), bounded samples, alloys, and dirty metals. The conventional geometric contribution to AHC is defined in the clean-metal limit only \cite{Nagaosa10}, while instead our geometrical marker, when applied to a dirty metal, includes extrinsic contributions of geometrical nature as well.

Work supported by the ONR Grant No. N00014-12-1-1041.


\begin{thebibliography}{10}
\bibitem{sign}
{ The choice of the sign in $C_1$ is not uniform across the literature; the
  choice made here agrees with most of the cited papers}.
\end{thebibliography}
\end{document}